\documentclass[conference]{IEEEtran}
\usepackage{cite}
\usepackage{graphicx}
\usepackage{tikz}
\usetikzlibrary{shapes,backgrounds}
\usetikzlibrary{positioning}
\usepackage[top=.81in, bottom=.81in, left=.81in, right=.81in]{geometry}  
\usepackage{algorithm} 
\usepackage[absolute]{textpos}
\usepackage{algorithmic}
\usepackage{multirow}
\usepackage{threeparttable}
\usepackage[strict]{changepage}
\usepackage[keeplastbox]{flushend} 
\usepackage{bm,upgreek} 
\usepackage{amssymb}
\usepackage{bbm}
\usepackage{hhline}
\usepackage{enumitem} 
\ifCLASSINFOpdf
\else
\fi
\usepackage{amsmath}
\ifCLASSOPTIONcompsoc
  \usepackage[caption=false,font=normalsize,labelfont=sf,textfont=sf]{subfig}
\else
  \usepackage[caption=false,font=footnotesize]{subfig}
\fi
\begin{document}
%
\title{Partially Blind Handovers for mmWave\\New Radio~Aided~by~Sub\nobreakdash-6~GHz~LTE~Signaling}
%
%
%

\author{\IEEEauthorblockN{Faris B.~Mismar and Brian L.~Evans}

\IEEEauthorblockA{Wireless Networking and Communications Group, The University of Texas at Austin, Austin, TX 78712 USA}
}

\maketitle

\begin{abstract}
For a base station that supports cellular communications in sub-6 GHz LTE and millimeter (mmWave) bands, we propose a supervised machine learning algorithm to improve the success rate in the handover between the two radio frequencies using  sub-6 GHz and mmWave prior channel measurements within a temporal window.  The main contributions of our paper are to 1) introduce partially blind handovers, 2) employ machine learning to perform handover success predictions from sub-6 GHz to mmWave frequencies, and 3) show that this machine learning based algorithm combined with partially blind handovers can improve the handover success rate in a realistic network setup of colocated cells. Simulation results show improvement in  handover success rates for our proposed algorithm compared to standard handover algorithms.

\end{abstract}


%
\IEEEpeerreviewmaketitle

\section{Introduction}
\begin{figure}[!t]
\centering
\begin{tikzpicture}[node distance = 4em, font=\scriptsize]
    \node [rectangle, draw, 
    text width=8em, text centered, rounded corners, minimum height=1em, fill=gray!30] (SON) {SON};
    \node [rectangle, draw, 
    text width=8em, text centered, rounded corners, minimum height=1em, below of=SON, yshift=-1.3cm] (mmWave) {mmWave Measurements};
    \path [draw, -latex] (-2.5,0)  -- node [near start, xshift=-0.4cm, left, text width=7em] {Acceptance threshold $\varepsilon$, Collection period $T$} (SON.180);
  \path [draw, latex-latex,postaction={draw, line width=0.05cm},color=gray!50] (SON.330) -- node {} (mmWave.30);
    \node [rectangle, draw, fill=white,
    text width=8em, text centered, rounded corners, minimum height=1em, above of=mmWave] (sub6) {Sub-6 GHz LTE Measurements};
 \path [draw, latex-latex,postaction={draw, line width=0.05cm},color=gray!50] (SON.210) -- node {} (sub6.136);

     \draw (-1.4,-2.7) ellipse (3.2cm and 0.7cm) node [below left,yshift=-0.8cm,xshift=-0.6cm] {Radio Network (LTE and NR)};


      
\end{tikzpicture}
\caption{The machine learning based algorithm and partially blind handovers interacting with the self-organizing network (SON).}
\label{fig:overall}
\end{figure}
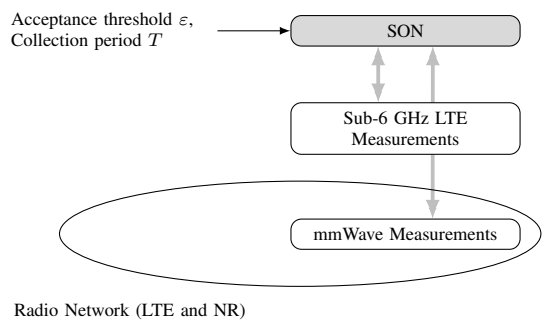

%
%
%
%
\IEEEPARstart{T}{he} fifth-generation (5G) of wireless communications, also known as \textit{New Radio} (NR), is believed to adopt the \textit{millimeter wave} (mmWave) frequencies in addition to sub-6 GHz band while other technologies such as LTE or \mbox{LTE-A} Pro (collectively LTE from now on) will continue to use frequencies in the sub-6 GHz band \cite{6515173}.  As a result, the need for measurement gaps prior to executing the \textit{inter-radio access technology} (inter-RAT) handover from these technologies to 5G. The major problem with measurement gaps is their reduction of the perceived end-user throughput because data transmission ceases during these gaps  \cite{3gpp36331}.  In this paper, we propose a machine learning based algorithm residing in the LTE base station (eNodeB) prior to configuring the measurement gap.  This step is essential before the 5G base station establishes the radio link into NR as part of the handover execution.  Such a step is important because the cost of a failed handover during a measurement gap is expensive: it can cause a dropped call, an increased packet delay, or low data throughput.  All of these are factors that contribute to the dissatisfaction of the end-user about the service or about the mobile network operator as a whole.  
Our approach learns a statistical relationship between both  sub-6 GHz in LTE and mmWave frequency measurements in NR and uses it to decide whether a handover would succeed if executed. This is an enabler for \textit{self-organizing networks} (SON) to perform as promised in 5G \cite{6963801}.  Fig.~\ref{fig:overall} shows the overall setup.



 

During measurement gaps, the data transmission ceases so the \textit{user equipment} (UE) receiver circuitry can measure the different frequency band in which the target technology typically operates.  In our case, the target technology that the LTE-connected UE wants to measure is the 5G technology operating in the mmWave frequency range.



In \cite{7925685}, the authors used \textit{correlation based adaptive compressed sensing} (CBACS) in mmWave to estimate with high-accuracy the \textit{channel state information} (CSI) particularly the \textit{angle of departure} (AoD) and \textit{angle of arrival} (AoA), which are essential for hybrid beamforming.  Compressed sensing is justified since mmWave radio frequency waves scatter poorly as they propagate.  Their approach decides the CSI by comparing correlations of the received CSI versus the quantized sensing vectors.  They obtained higher accuracy without an increase in the training overhead.  We occasionally override the need for CSI by allowing the base station to make the channel estimation on behalf of the UE based on collected~data.

Using \textit{orthogonal matching pursuit} (OMP), the authors in \cite{7458188} formulated a sparse signal recovery problem, which exploits the sparse nature of mmWave channels.  They performed channel estimation based on a parametric channel model with quantized AoAs and AoDs.  They bounded their results by the guaranteed OMP performance as an upper bound and the oracle estimator which assumes perfect knowledge of the AoAs and AoDs.  They asserted that estimating the CSI in mmWave with massive MIMO systems is challenging because the signal-to-noise ratio (SNR) before beamforming is low and the number of antennas is generally high.  This is an important finding that we overcome in our paper by using measurements from the sub-6 GHz system to estimate the received signal powers in the mmWave band.

The authors in \cite{8198818} attempted to exploit the use of out-of-band information to estimate the mmWave signal powers.  They proposed using three categories of out-of-band information: 1) position information such as global positioning system (GPS) signals, cameras, or even radars, 2) sub-6 GHz signals coming from Wi-Fi or cellular deployments, and 3) wireless sensor networks.  The authors did not consider the use of statistical learning or the reuse of the mmWave measurements for the same users within a temporal window, which we are proposing.  Further, the paper does not cover the various radio protocols of the technologies they referenced (e.g., the handover procedure).

Handovers in 5G wireless systems are expected to resemble what LTE offers at the \textit{radio resource control} (RRC) layer \cite{3gpp36331}.  These handovers can be broadly grouped into:
\begin{itemize}
\item\textit{Measured:} the UE measures the source cell and, in specific cases, the target cell radio conditions and reports its measurement as an event to the base station.
\item\textit{Blind:} the UE has performed an action (other than a \mbox{UE-reported} radio measurement, such as requesting a service) which triggers the base station to reconfigure the radio bearers to a new cell offering this service.  This cell can be in a different frequency or a different technology. 
\end{itemize}

Estimating the mmWave frequencies from the \textit{colocated} LTE serving cell measurements in the sub-6 GHz band helps in preempting the handover procedure if it is likely to fail due to weak mmWave signal levels.  This is done without having the UE explicitly measure the mmWave carrier prior to the handover.  We call this handover \textit{``partially blind"} because it still requires target technologies measurement reports for \textit{both} technologies for a certain period of time $T$ (the \textit{collection period}) to build the training data required for the machine learning algorithm to operate.  It is blind in that the measurement gap configuration is temporarily unnecessary after the collection period has passed (details in Section~\ref{sec:system_model}).  Owed to the partially blind handovers, the UE data rates are not reduced due to a superfluous measurement gap.  We have chosen LTE sub-6 GHz and mmWave bands for this paper; however we believe that this approach works for any inter-frequency pairs provided the cells are colocated and their radio environment parameters are known.  For the particular case of mmWave and sub-6 GHz frequencies in this paper, since the mmWave wavelengths shrink by an order of magnitude relative to sub-6 GHz frequencies, the material penetration incurs greater attenuation as stated in \cite{7999294}.  This therefore elevates the importance of line-of-sight (LOS) propagation in both bands, with a difference that the the path loss model in mmWave frequencies \cite{7999294, 7481506} is a random blocking model unlike the model in the sub-6 GHz frequencies.

Our main contributions in this paper are as follows:
\begin{enumerate}
\item Introduce the new concept of partially blind handovers.
\item Employ machine learning to perform handover success predictions from sub-6 GHz to mmWave frequencies.
\item Show that this machine learning based algorithm combined with partially blind handovers can improve the handover success rate in a realistic network setup of colocated cells of both frequencies.
\end{enumerate}

\section{System Model}\label{sec:system_model}
This system comprises two components:
\begin{enumerate}
\item A network of two colocated cells and connected users in an outdoor setting in a dense-urban environment.
\item A machine learning algorithm using \textit{extreme gradient boosting} (XGBoost) classifier to override the handover decision if needed based on the estimation inferred using the history of the handover success for that user.
\end{enumerate}

For the model to be valid, the collection period $T$ cannot exceed the \textit{channel coherence time}.  This is the temporal window during which measurements are collected.  As not all the UEs require handovers, the number of data points collected cannot exceed the number of handover attempts.  The eNodeB reruns the algorithm for a given UE every time the UE establishes a radio connection or hands off to a new eNodeB.

\subsection{Radio Network}\label{sec:radio}

The radio network is comprised of two colocated cells (i.e., an overlay) with a circular geometry with a radius $r$ in a dense urban setting.  Each cell operates a different technology and frequency band.

Similar to any cellular network today, the UEs measure the downlink radio frequencies and report them to the base stations.  The difference here is that the base stations can decide whether to configure a measurement gap in LTE, since they now know by means of our machine learning based algorithm that the 5G mmWave band will not have sufficient signal power levels to maintain a session in that band.  Further, we have handovers to facilitate service continuation as the UE reaches to edge of coverage.

We distribute the UEs in  the network according to a \textit{homogeneous poisson point process} (PPP)  \cite{bacelli}.  The process $\Phi$ has an \textit{intensity parameter} $\lambda$ representing the expected number of users served per unit area.  We define the point process $\Phi$ for the number of users $N$ in the network service area $W$.  The UEs are sampled from a Poisson distribution with mean $\lambda W \triangleq \lambda \pi r^2$, where $r$ is the cell radius.

\begin{figure}[!t]
\centering
\begin{adjustwidth}{1.2cm}{0cm}
\includegraphics[scale=0.45]{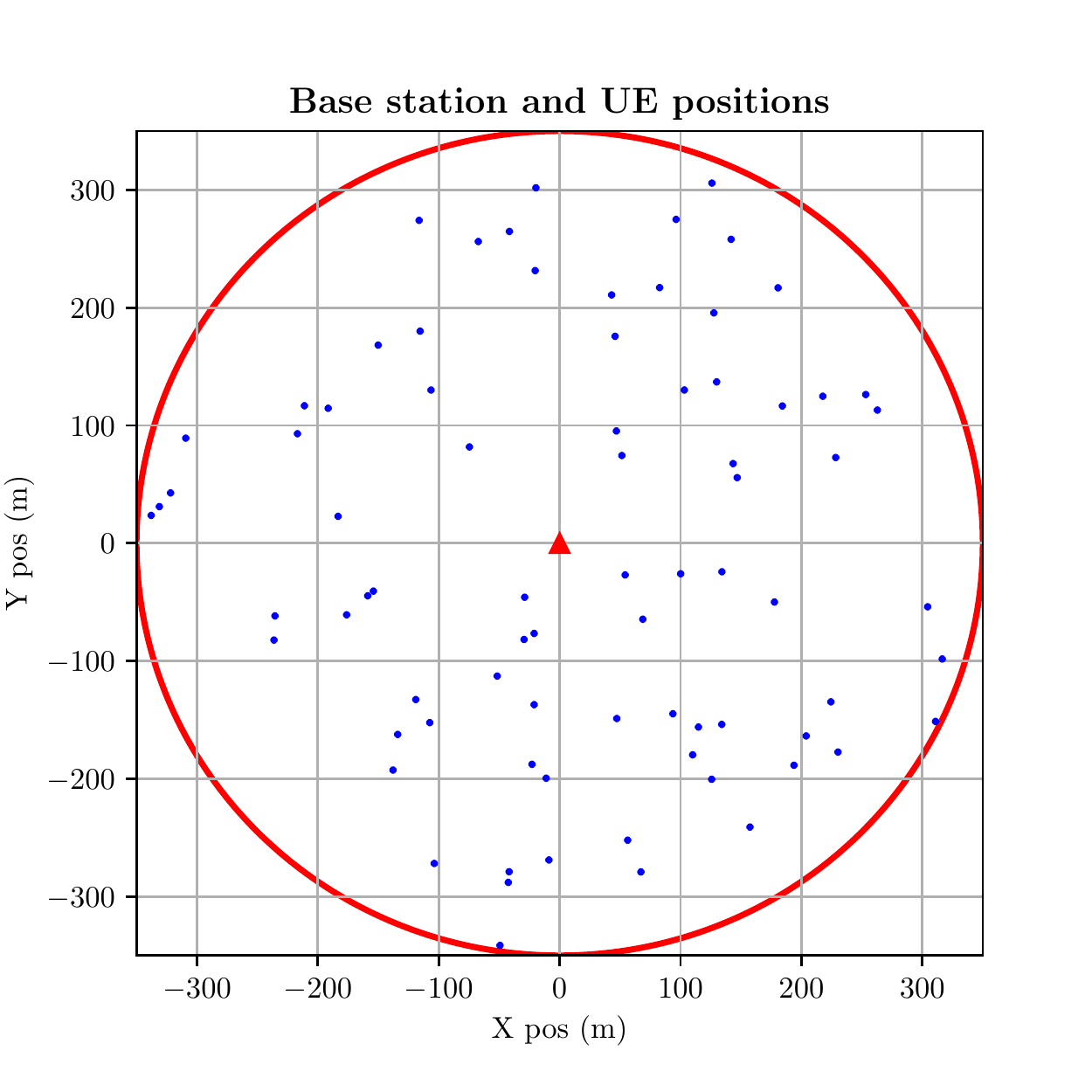}
\end{adjustwidth}
\vspace*{-0.1in}
\caption{The network layout simulated at $t = 0$. UEs are blue dots and the base station is the red triangle in the origin.  The blue dots move throughout $1\le t \le T_\text{sim}$ by resampling from the poisson point process.}
\label{fig:network}
\end{figure}

The $i$-th UE position is i.i.d.~sampled from a continuous uniform distrubtion in $\mathbb{R}^2$ using the polar coordinates $(r_i, \theta_i)$ where $0 \le r_i \le r$, $0 \le \theta_i \le 2\pi$, and $i = 1, 2, 3, \ldots, N$. Fig.~\ref{fig:network} shows the layout of the PPP in the serving area of the colocated cells just before simulation starts.  The radio parameters of this network are in Table~\ref{table:rf_parameters}. 

\subsection{Machine Learning}
We have chosen the XGBoost classifier for our predictions because it can perform parallel computing on trees hence making the proposed algorithm suitable for distributed base stations.  It is also invariant to input scaling and can learn higher order interaction between features. XGBoost is a scalable ensemble learning technique discussed in \cite{Chen:2016:XST:2939672.2939785}.  

XGBoost minimizes an objective function which has a differentiable convex loss function and regularization terms $\alpha\Vert \mathbf{w} \Vert_1 + \frac{1}{2}\lambda\Vert \mathbf{w}\Vert^2_2 + \gamma T$, where $\mathbf{w}$ is the vector containing the leaf weights in the boosted tree and $T$ is the number of leaves. Regularization controls the complexity of the model and therefore helps avoid overfitting.

The $m\times n$ matrix of learning features $\mathbf{X} \triangleq [\mathbf{x_i}]_{i = 1}^n$ is listed in Table~\ref{tab:features} where $n$ is the number of features and each feature $\mathbf{x_i}$ is an $m$-dimensional vector obtained during the measurement collection period.  The supervisory label vector is $\mathbf{y}$.
\begin{table}[!t]
\setlength\doublerulesep{0.5pt}
\caption{Machine Learning Features $\mathbf{X}$}
\label{tab:features}
\centering
\begin{adjustwidth}{-0.3cm}{0cm}
\begin{tabular}{ l|lll } 
\hhline{====}
& Parameter & Type & Description \\
 \hline
$\mathbf{x_1}$ & $(x, y)$ & Float & Coordinates of UE. \\
$\mathbf{x_2}$ & Distance & Float & The Euclidean distance from the base \\
& & & station based on the coordinates of UE. \\
$\mathbf{x_3}$ & \texttt{RSRP\_}x & Float & RSRP in x = \{LTE, mmWave\} bands.\\
$\mathbf{x_4}$ & \texttt{Gap\_Closed} & Boolean & Did UE report event A1 based on \\
& &  & its RSRP measurement?\\
$\mathbf{x_5}$ & \texttt{Gap\_Open} &  Boolean & Did UE report event A2 based on \\
& &  & its RSRP measurement?\\
\hhline{====}
\end{tabular}
\end{adjustwidth}
\end{table}
The supervisory label is an integer with 0 being handover not executed and 1 being executed.

The features $\mathbf{x_3}$, $\mathbf{x_4}$, and $\mathbf{x_5}$ are obtained directly from UE measurements.  However, features $\mathbf{x_1}$ and $\mathbf{x_2}$ require additional modification to the standards to enable the UEs to report their coordinates through RRC messaging.  These coordinates can come from the positioning methods such as \textit{global navigation satellite systems} (GNSS) or \textit{observed time difference of arrival} (OTDOA) as defined in the LTE positioning protocol \cite{3gpp36355}.

We tune the hyperparameters in Table~\ref{table:ml_parameters} using grid search on $K$-fold cross-validation.  Cross-validation is required to tune the model for accuracy.

The time complexity of our algorithm is $O\left(mn \cdot (d_\text{max} E + n \log n)\right)$ using \cite{Chen:2016:XST:2939672.2939785} where $d_\text{max}$ is the maximum depth of the boosted tree and $E$ is the total number of trees or estimators.  The complexity is a function of the number of UEs covered in the cell and the measurement reporting frequency, but not the number of cells in the network.

\section{Handover Algorithms}
\subsection{Baseline Inter-RAT Handover Algorithm}\label{sec:baseline}
\begin{figure}[!t]
\centering
\begin{tikzpicture}[node distance=3cm, auto]
\tikzstyle{every node}=[font=\small]

\node[draw, rectangle] (UE)  at (0,0) {UE};
\node[draw, rectangle, right of=UE] (eNB) {eNodeB};
\node[draw, rectangle, right of=eNB] (NR) {5G Base Station};
\draw (0,-0.25) -- (0,-5.5);
\draw (3,-0.25) -- (3,-5.5);
\draw (6,-0.25) -- (6,-5.5);

\draw[->, >=latex] (0,-1)  -- node [text width=4cm, align=center, above] {\scriptsize{RRC Measurement Report: \\ Event A2}} ++(3,0);
\node[circle,fill=black,inner sep=0pt,minimum size=5pt,label=right:{A}] (a) at (3,-1) {};
\draw[<-, >=latex, dashed, black!50] (0,-2)  -- node [text width=4cm, align=center, midway,above] {\scriptsize{RRC Connection Reconfiguration: \\ Measurement gap configuration}} ++(3,0);
\draw[<->, dashed, black!50,>=latex, line width=0.35mm] (0,-3)  -- node [text width=4cm, align=center, above] {\scriptsize{Measure the mmWave power: \\  Above minimum HO threshold}} ++(6,0);
\draw[->, dashed, black!50, >=latex] (0,-4)  -- node [text width=4cm, align=center, midway,above] {\scriptsize{RRC Measurement Report:\\ Event B2}} ++(3,0);
\node[rectangle,fill=black,inner sep=0pt,minimum size=5pt,label=right:{D}] (d) at (3,-4) {};
\draw[<-, >=latex ] (0,-5)  -- node [text width=4cm, align=center, midway,above] {\scriptsize{RRC Mobility From E-UTRA Command}} ++(3,0);
\node[circle,fill=black,inner sep=0pt,minimum size=5pt,label=right:{B}] (b) at (3,-5) {};
\draw[<->, >=latex] (0,-5.5)  -- node [text width=4cm, align=center, midway,above] {\scriptsize{Random Access}} ++(6,0);
\end{tikzpicture}
\caption{Handover signaling procedure adapted after \cite{3gpp36331}. The handover decision point D and the metric trigger points A and B are also shown. (RRC stands for \textit{radio resource control}, HO is short for handover, E-UTRA is the \textit{evolved universal terrestrial radio access}, eNodeB is the LTE base station). Signals in dashed gray are ones impacted by the proposed algorithm.}
\label{fig:ho_proc}
\end{figure}
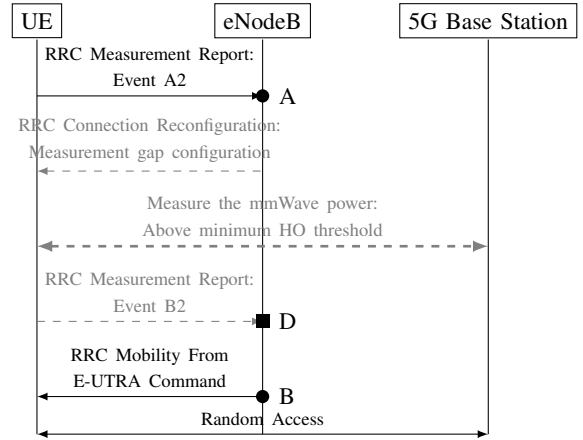

Industry standards \cite{3gpp36331} specify that when the UE measures the \textit{reference symbol received power} (RSRP) of a cell which is worse than a threshold, it triggers RRC event A2. This allows inter-RAT measurements to start for this UE through measurement gaps. Also, when the UE measures an RSRP better than a threshold, it triggers RRC event A1  ending inter-RAT measurements for this UE. If mmWave power is above a certain threshold, UE triggers RRC event B2 and proceeds with random access towards the mmWave carrier. The handover is then successfully executed.    Fig.~\ref{fig:ho_proc} shows the baseline procedure through which a handover is prepared and executed. Trigger point A is where the handover attempt is stepped and trigger point B is where the handover execution counter is stepped after eNodeB decides to allow handover. 

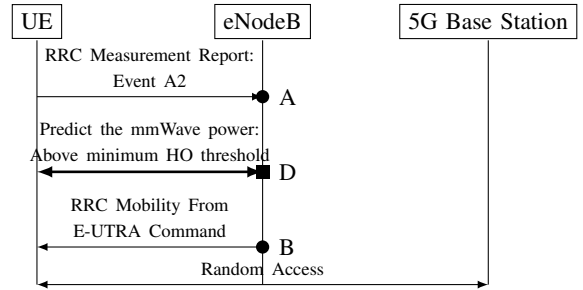
\begin{figure}[!t]
\centering
\begin{tikzpicture}[node distance=3cm, auto]
\tikzstyle{every node}=[font=\small]

\node[draw, rectangle] (UE)  at (0,0) {UE};
\node[draw, rectangle, right of=UE] (eNB) {eNodeB};
\node[draw, rectangle, right of=eNB] (NR) {5G Base Station};
\draw (0,-0.25) -- (0,-3.5);
\draw (3,-0.25) -- (3,-3.5);
\draw (6,-0.25) -- (6,-3.5);

\draw[->,>=latex] (0,-1)  -- node [text width=4cm, align=center, above] {\scriptsize{RRC Measurement Report: \\ Event A2}} ++(3,0);
\node[circle,fill=black,inner sep=0pt,minimum size=5pt,label=right:{A}] (a) at (3,-1) {};
\draw[<->, >=latex, line width=0.35mm] (0,-2)  -- node [text width=4cm, align=center, above] {\scriptsize{Predict the mmWave power: \\ Above minimum HO threshold }} ++(3,0);
\node[rectangle,fill=black,inner sep=0pt,minimum size=5pt,label=right:{D}] (d) at (3,-2) {};
\draw[<-, >=latex] (0,-3)  -- node [text width=4cm, align=center, midway,above] {\scriptsize{RRC Mobility From E-UTRA Command}} ++(3,0);
\draw[<->, >=latex] (0,-3.5)  -- node [text width=4cm, align=center, midway,above] {\scriptsize{Random Access}} ++(6,0);
\node[circle,fill=black,inner sep=0pt,minimum size=5pt,label=right:{B}] (b) at (3,-3) {};
\end{tikzpicture}
\caption{Proposed handover signaling procedure changing the decision point D earlier in the procedure and allowing the eNodeB to predict the mmWave band measurement.}
\label{fig:ho_proc_proposed}
\end{figure}


\subsection{Proposed Inter-RAT Handover Algorithm}\label{sec:algorithm}
In the proposed Algorithm~\ref{alg:the_alg}, the decision of whether to accept the UE measurement or override it is based on \textit{receiver operating characteristic} area under the curve (ROC AUC) of true positive rate vs false positive rate.  This curve is obtained from the machine learning technique which uses training and cross-validation data to predict whether handover will succeed or fail.  Then if the LTE received power is worse than the same threshold in the baseline algorithm and the \textit{predicted} mmWave received power is above a certain threshold or if the ROC AUC is inadequate, the algorithm proceeds as the baseline. Alternatively, if the predicted mmWave received power is lower than the threshold, the eNodeB will preempt the request from the UE to handover to mmWave thereby saving the UE from a handover likely to fail. Fig.~\ref{fig:ho_proc_proposed} shows that the number of steps in the proposed procedure has been reduced due to the machine learning prediction.

 \begin{algorithm}[!t]
 \small
\caption{\small Partially blind handover success estimation}
\label{alg:the_alg}
\begin{algorithmic}[1]
\renewcommand{\algorithmicrequire}{\textbf{Input:}}
\renewcommand{\algorithmicensure}{\textbf{Output:}}
 \REQUIRE Parameters listed in Table~\ref{table:ml_parameters} and Table~\ref{table:rf_parameters}, $\varepsilon$ the acceptance threshold, and the simulation time $T_\text{sim}$.
 \ENSURE An output table for all UEs containing a time sequence showing whether the handover to 5G must be overridden or not based on estimated mmWave received signal level.

 \STATE Compute $N$ the total number of UEs in the cell per Section~\ref{sec:radio}.
  \FOR{$i\in\{1,...,N\}$} 
  \STATE Obtain the generated simulation data for UE $i$ for all times $t = 1, \ldots, T_\text{sim}$, which are the features $\mathbf{X}$ listed in Table~\ref{tab:features}.
  \STATE Compute the handover success of this UE: measurement gap opened AND mmWave power greater than threshold, which is the supervisory label $[\mathbf{y}]_{i}$ for UE $i$.
  \STATE Split UE $i$ data to training data and test data.  Training data is collected over a period $1, \ldots, T$, where $T \triangleq \min(T_\text{coherence}, \lceil r_\text{training}\cdot T_\text{sim}\rceil)$.
  \STATE Train the XGBoost model using the training data and use grid search on $K$-fold cross-validation to tune the hyperparameters.
  \STATE Using the trained model obtain the proposed handover execution decision $\mathbf{\hat{y}}$.
  \STATE Obtain the area under the ROC curve for this model (ROC AUC).
  \IF {(ROC AUC $\ge \varepsilon $)}
  \STATE Use $\hat{y}$ as a valid estimate of handover execution decision (follow Fig.~\ref{fig:ho_proc_proposed}).
  \ELSE 
   \STATE Use the UE reported measurements (baseline algorithm).
   \ENDIF
  \ENDFOR 
\end{algorithmic}
\end{algorithm}

\section{Simulation Results}\label{sec:simulation}
\begin{figure}[!t]
\begin{adjustwidth}{-1cm}{0cm}
\centering
\includegraphics[width=0.5\textwidth]{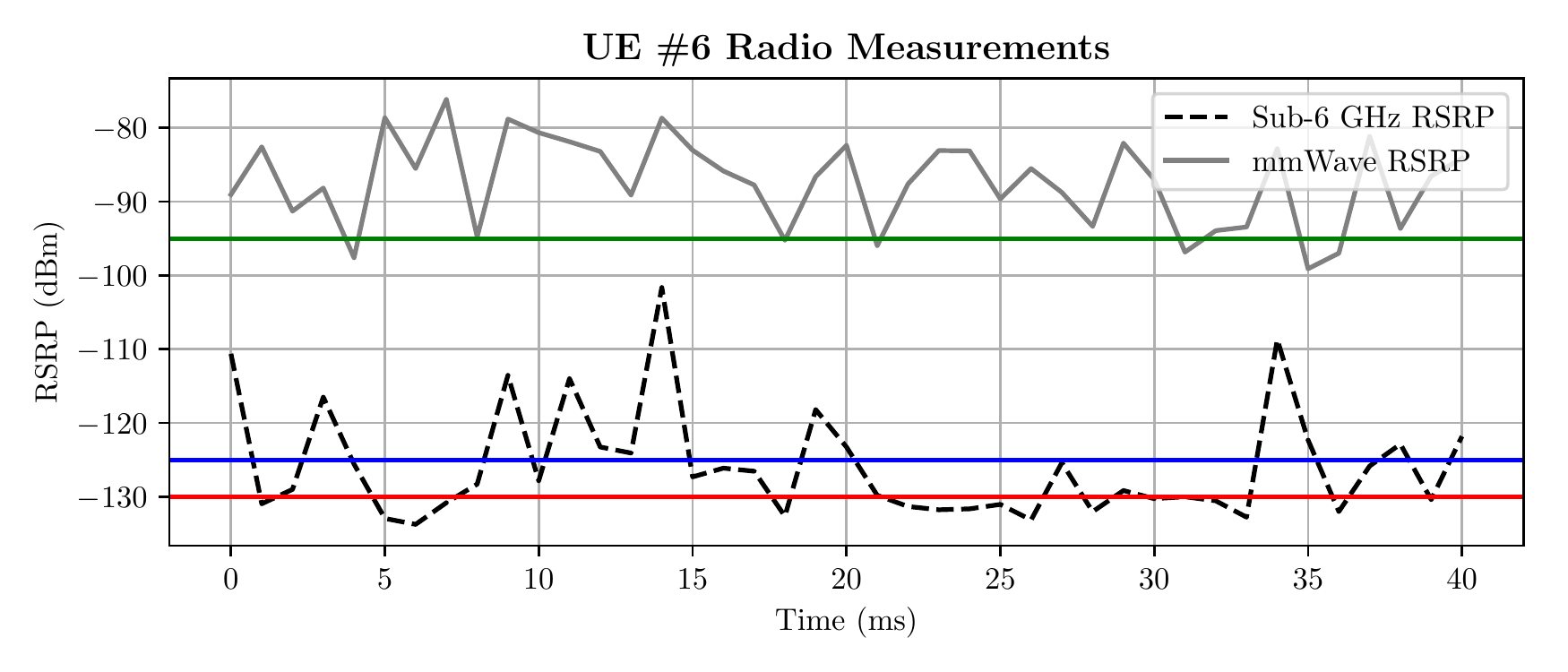}
\end{adjustwidth}
\vspace*{-0.2in}
\caption{Simulated received power for UE \#6.  The blue, red, and green lines are the RRC events (A1, A2, and B2) thresholds respectively.}
\label{fig:simulation1}
\end{figure}

\begin{figure}[!t]
\begin{adjustwidth}{-0.7cm}{0cm}
\centering
\includegraphics[width=0.51\textwidth]{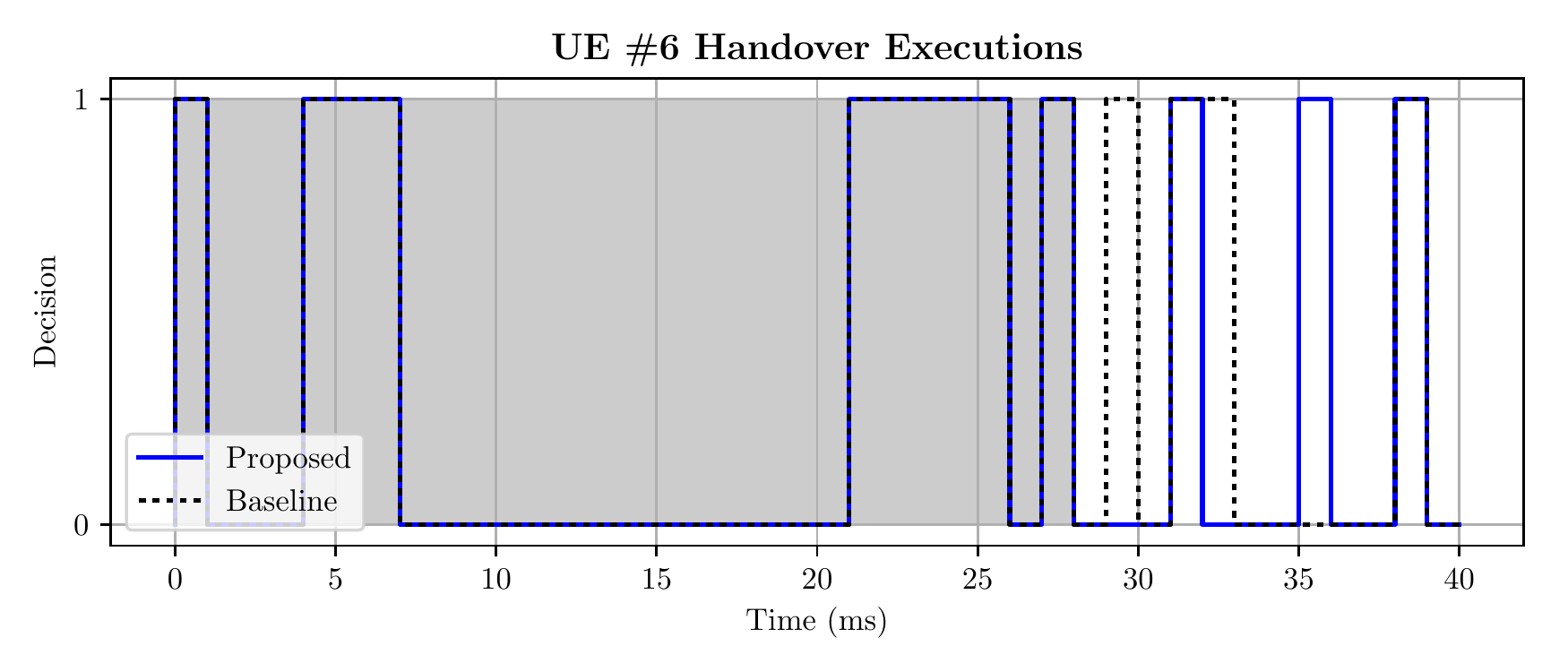}
\end{adjustwidth}
\vspace*{-0.2in}
\caption{Handover executions for UE \#6 over the simulation time.  The dotted black line denotes the baseline and the solid blue line is the proposed algorithm.  The shaded region is the region corresponding to the collection period $T$ in which the handover execution decisions are identical and excluded from results.  Outside the shaded region the handover execution decisions differ.}

\label{fig:simulation2}
\end{figure}

\begin{figure}[!t]
\centering
\includegraphics[width=0.4\textwidth]{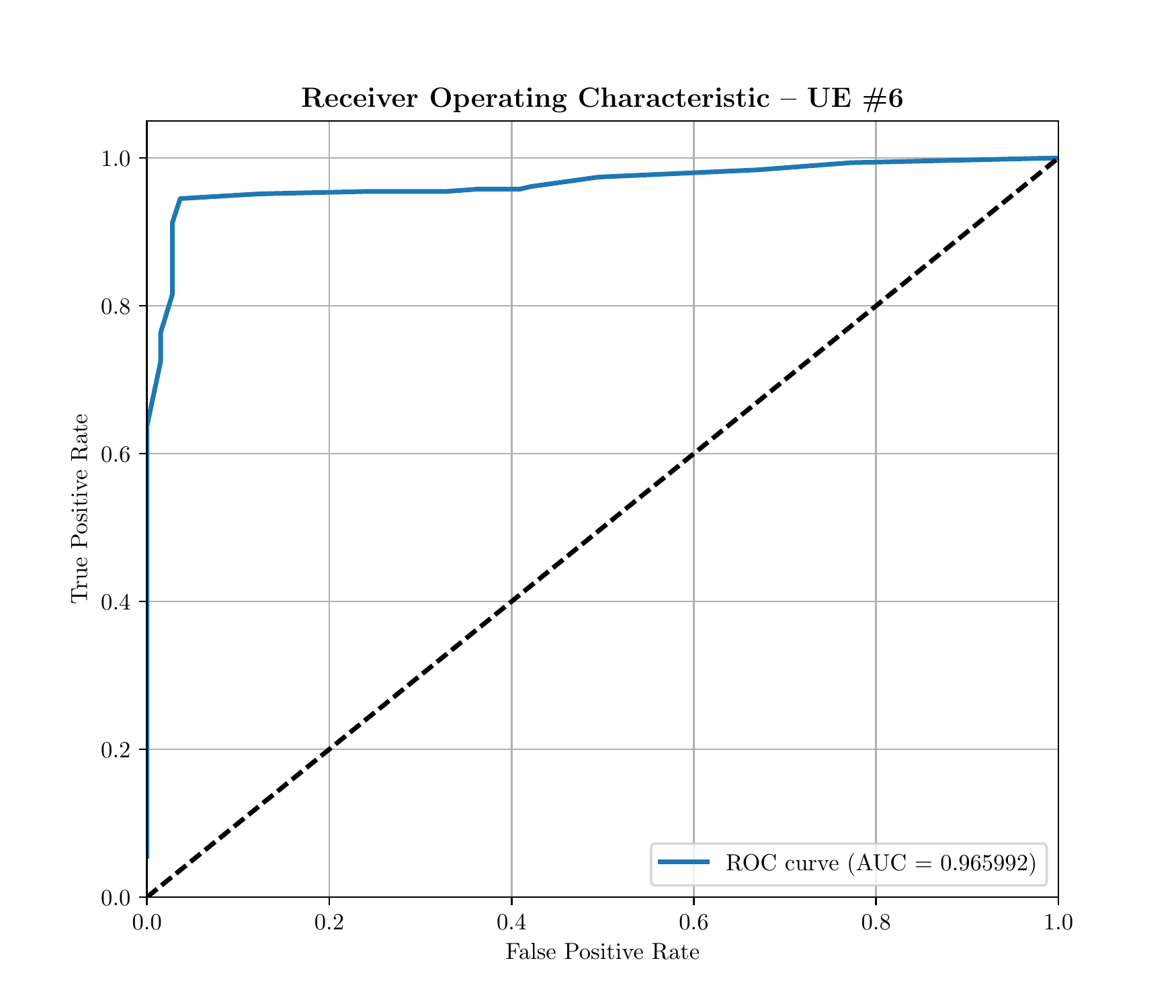}
\vspace*{-.2in}\caption{Receiver operating characteristic (ROC) curve for UE \#6 using $T_\text{sim} = 800$ ms and $\lambda = 2\times10^{-4}$.}
\label{fig:roc}
\end{figure}

The performance of our algorithm has been tested using using a computer simulation written in Python on GitHub \cite{mycode}.   The machine learning hyperparameters and the radio environment parameters are in Tables~\ref{table:ml_parameters} and \ref{table:rf_parameters} respectively.  We showed graphical outputs of an arbitrarily selected UE.  Fig.~\ref{fig:simulation1} shows the RSRP for the same UE over the simulation time, for signals in both sub-6 GHz and mmWave frequencies, with the RRC event thresholds shown as horizontal lines.  Fig.~\ref{fig:simulation2} shows the corresponding handover execution of both algorithms versus simulation time where the handover decisions are identical for a period equal to 28~ms ($= 0.7\times 40$ ms).  The machine learning algorithm has then gathered enough information to be able to predict if an upcoming handover is likely to fail or succeed.  It therefore generates handover decisions different from the baseline.  Finally, Fig.~\ref{fig:roc} shows the ROC curve for the same UE, where the algorithm performance measured by the area under this curve (ROC AUC) has passed the $\varepsilon = 0.7$ threshold.  This area is a measure of the ability of our algorithm to predict an IRAT handover success.  This area ranges between 0.5 and 1 and is represented by $\varepsilon$ in Algorithm~\ref{alg:the_alg}.  While the threshold at which the area deems a prediction valid in our algorithm is arbitrary, an area that is equal to 0.5 means the model is providing random guesses and cannot be used.


\begin{table}[!t]
\setlength\doublerulesep{0.5pt}
\caption{Machine Learning Hyperparameters for XGBoost}
\label{table:ml_parameters}
\centering
\begin{tabular}{ lr } 
\hhline{==}
Parameter & Value \\
 \hline
Training data split $r_\text{training}$  & 0.7\\
Objective & \{logistic, linear\}\\
$K$-fold crossvalidation $K$ & 5 \\
Number of estimators $E$ & 500 \\
$\ell_1$ regularization term $\alpha$& \{0, 0.5, 1\}\\ 
$\ell_2$ regularization term $\lambda$ & \{0, 0.5, 1\} \\
Complexity control $\gamma$ &  \{0, 0.02\} \\
Sample weights & \{0.5, 0.7\} \\
Child weights & \{0, 1, 10\} \\
Max depth $d_\text{max}$ & \{6, 8\} \\
\hhline{==}
\end{tabular}
\end{table}

\begin{table}[!t]
\setlength\doublerulesep{0.5pt}
\vspace*{-0.1in}
\caption{Radio Environment Parameters}
\label{table:rf_parameters}
\centering
\begin{threeparttable}
\begin{tabular}{ lr } 
\hhline{==}
Parameter & Value \\
 \hline
LTE bandwidth & 20 MHz \\ 
LTE center frequency & 2.1 GHz \\
LTE cyclic prefix & normal \\
5G mmWave bandwidth & 100 MHz \\
5G mmWave center frequency & 28 GHz \\
LTE Propagation model & COST 231 (LOS; no shadowing) \\
5G mmWave Propagation model & \cite{7481506} \\
Morphology & urban \\
PPP intensity parameter $\lambda$ & $2\times \{10^{-5}, 10^{-4}\}$\\
Simulation time $T_\text{sim}$ & \{40, 400, 800\} ms \\
Cell radius $r$ & 350 m \\
LTE base station power & 46 dBm \\
5G base station power & 46 dBm \\
Antenna pattern & omnidirectional\tnote{\textdagger} \\ 
Antenna height & 20 m \\
Antenna sub-6 GHz gain & 17 dBi \\
Antenna mmWave gain & 24 dBi \\
Cellular geometry & circular \\
UE height & 1.5 m \\
RRC event A1 trigger & -125 dBm \\
RRC event A2 trigger & -130 dBm \\
RRC event B2 trigger &  -95 dBm \\
RRC events time-to-trigger & 0 sec. \\
\hhline{==}
\end{tabular}
\begin{tablenotes}
\item  \footnotesize \textdagger Synthesized omnidirectional pattern using directional antennas.
\end{tablenotes}
\vspace*{-0.1in}
\end{threeparttable}

\end{table}

\begin{table*}[!t]
\setlength\doublerulesep{0.5pt}
\begin{adjustwidth}{-0.4cm}{0cm}
\caption{Handover success rate simulation results}
\label{table:results_1}
\centering
\begin{tabular}{c|c|c|cccc} 
\hhline{======}
Sim. Time & Rate & Algorithm & Attempts & Failures & Success rate \\
\hline
\multirow{3}{*}{$T_\text{sim} = 40$ ms} & $\lambda = 2\times 10^{-4} $ &Baseline & 1,259 & 58 & 95.39\%\\
&  $(N = 78)$ & Proposed & 1,259 & \textbf{55} & \textbf{95.63\%} \\
\cline{2-6}
& $\lambda = 2\times 10^{-5}$  & Baseline & 135 & 6 & 95.56\%\\
& $(N = 8)$ & Proposed &  135  &  6 & \textbf{95.56\%} \\
\hline
\multirow{3}{*}{$T_\text{sim} = 400$ ms}  &$\lambda = 2\times 10^{-4}$ & Baseline & 12,601 & 695 & 94.48\% \\
& $(N = 78)$ & Proposed & 12,601 & \textbf{450} & \textbf{96.43\%} \\
\cline{2-6}
& $\lambda =2\times 10^{-5}$ & Baseline & 1,254  &  70 & 94.42\%\\
& $(N =8)$ & Proposed & 1,254  & \textbf{13} & \textbf{98.96\%} \\
\hline
\multirow{3}{*}{$T_\text{sim} = 800$ ms} & $\lambda = 2\times 10^{-4}$  & Baseline & 25,070  & 1,378 & 94.50\% \\
& $(N =78)$ & Proposed &25,070  & \textbf{757} & \textbf{96.98\%} \\
\cline{2-6}
& $\lambda = 2\times 10^{-5}$ & Baseline &  2,535 & 132 &94.79\%\\
&$(N = 8)$ & Proposed &  2,535 & \textbf{19} & \textbf{99.25\%} \\
\hhline{======}
\end{tabular}
\end{adjustwidth}
\end{table*}

Table~\ref{table:results_1} shows the results of the simulation.  We ran the simulation for three different durations ranging from short to long.  The coherence time of the channel falls in the range.    These different time durations and different number of UEs in the cell also show the robustness of the algorithm.  The improved numbers are in \textbf{boldface}.

With the least number of users and shortest duration, we see that the number of failures in both algorithms is similar. However, as the number of users or the simulation time increases, the number of failures by the proposed algorithm are improved.  Empirically, we observe that the baseline algorithm represents an upper bound of the number failures.
\vspace*{-0.2in}
\section{Conclusion}\label{sec:conclusion}
In this paper, we proposed the use of machine learning in making base stations predict handover success using both sub-6 GHz and mmWave prior measurements, effectively overriding the UE measurements using data collected within the coherence time.  The proposed contribution improved the \mbox{inter-RAT} handover success rate keeping sessions in the optimal band for longer amount of time.  If incorporated in 5G SON, this predictive algorithm can help facilitate the strong demand of high availability, high bandwidths, and low latencies promised by 5G by keeping away UEs likely to experience degraded service at the handover time through the machine learning based prediction algorithm we proposed.
\vspace*{-0.2in}

%








\ifCLASSOPTIONcaptionsoff
  \newpage
\fi


\bibliography{references}  
\bibliographystyle{IEEEtran}

\end{document}